\documentclass{svjour3mod}
\smartqed

\usepackage{graphicx}
\usepackage{amsmath}
\usepackage{amssymb}
\usepackage{bm}
\usepackage{verbatim}
\usepackage{algorithm2e}
\usepackage{multirow}
\usepackage{color}
\usepackage{url}
\usepackage{tikz}

\journalname{International Journal of Theoretical Physics}

\begin{document}
\usetikzlibrary{backgrounds,fit,decorations.pathreplacing,calc}

\newcommand{\ket}[1]{\ensuremath{\left|#1\right\rangle}}
\newcommand{\bra}[1]{\ensuremath{\left\langle#1\right|}}

\title{A Resilient Quantum Secret Sharing Scheme}
\author{Arpita Maitra \and Goutam Paul}
\institute{Arpita Maitra \at
        Applied Statistics Unit,
        Indian Statistical Institute, Kolkata 700 108, India\\
        \email{arpita76b@rediffmail.com}
\and
        Goutam Paul ({\em{Corresponding Author}}) \at
	Cryptology and Security Research Unit,
        R. C. Bose Centre for Cryptology and Security,
        Indian Statistical Institute, Kolkata 700 108, India\\
        \email{goutam.paul@isical.ac.in}
}


\maketitle
\begin{abstract}
A resilient secret sharing scheme is supposed to generate the secret correctly
even after some shares are damaged. In this paper, we show how quantum
error correcting codes can be exploited to design a resilient quantum
secret sharing scheme, where a quantum state is shared among more than one 
parties.



\CRclass{Quantum communication and cryptography}

\end{abstract}

\section{Introduction}
The quantum teleportation of a single qubit was first proposed by Bennett 
{\it et al.}~\cite {BB93}, using a maximally entangled bipartite quantum state. 
The idea of teleportation has been successfully applied in the process of 
quantum secret sharing. In quantum teleportation, Alice sends a qubit to a 
distant receiver Bob through some unitary operation involving the qubit and 
the entangled channel shared between them. Further results related to quantum
teleportation, that can be exploited in quantum secret sharing, include
results using multipartite quantum channels such as tripartite GHZ 
state~\cite{KB98}, four-partite GHZ state~\cite{P00}, an asymmetric W 
state~\cite{AP06} and the cluster state~\cite{BR01}. The perfect teleportation 
of an arbitrary two-qubit state was proposed using quantum channels formed by 
the tensor product of two Bell states~\cite{R05}, tensor product of two 
orthogonal states~\cite{YC06}, genuinely entangled five qubit state~\cite{P08},
five qubit cluster state~\cite{NL10} and six qubit genuinely entangled 
states~\cite{NJ10}. 

The idea of Quantum Secret Sharing (QSS) of a single qubit was first due to 
Hilery et al.~\cite{HBB99} using three and four qubit GHZ states. Later this 
process was investigated by Karlsson et al.~\cite{KK99} using three particle 
entanglement, Cleve et al.~\cite{C99} using a process similar to error 
correction and Zheng using W state~\cite{Z06}. The QSS of an arbitrary 
two-qubit state was proposed by Deng et al. using two GHZ states~\cite{D05}. 
QSS using cluster states was demonstrated by Nie~\cite{NS11},  
Panigrahi~\cite{PM11,PS11} and Han~\cite{H12}. Recently two qubit QSS 
was discussed using arbitrary pure or mixed resource states~\cite{ZL11} and 
asymmetric multipartite state~\cite{Z12}.

In this paper we do not exploit the ideas related to teleportation in 
quantum secret sharing. The idea of teleportation does not naturally take
care of the situation when some adversarial model is considered where some 
shares may be disturbed. Rather, we use quantum error correcting codes to
take care of the situation. There exist some works~\cite{rietjens05,zhang11}
for building quantum secret sharing schemes using (classical or quantum)
error correcting codes. However, none of these schemes addresses the resiliency
issue.

An arbitrary pure single-qubit quantum state is given by 
$\ket{\psi} = \alpha\ket{0} + \beta\ket{1}$ with 
$|\alpha|^2 + |\beta|^2 = 1$, where $\alpha, \beta \in \mathbb{C}$. In
quantum error correction scheme known as repetition code~\cite{nc04},
the above state is encoded as $\alpha\ket{000} + \beta\ket{111}$.
This encoding can correct a single bit-flip error. In~\cite{schindler13},
the authors pointed out that any measurement in the computational basis states
$\ket{0}$, $\ket{1}$ causes a projection onto the $\sigma_z$ axis of the
Bloch sphere and can be interpreted as an incoherent phase flip. Hence, any
protocol correcting against phase flips is sufficient to reverse measurements
in the computational basis. The repetition code can be modified to protect 
against such phase-flip errors by a simple basis change from 
$\ket{0}$, $\ket{1}$ to $\ket{\pm} = \frac{1}{\sqrt{2}}(\ket{0} \pm \ket{1})$. 

{\small
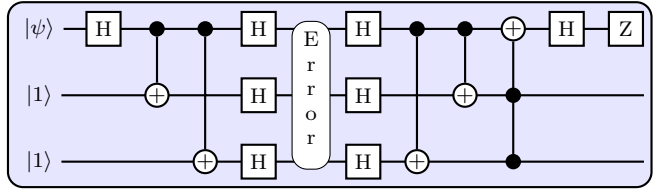
\begin{figure}
\centering
\begin{tikzpicture}[thick]
    \tikzstyle{operator} = [draw,fill=white,minimum size=1.5em] 
    \tikzstyle{phase} = [draw,fill,shape=circle,minimum size=5pt,inner sep=0pt]
    \tikzstyle{surround} = [fill=blue!10,thick,draw=black,rounded corners=2mm]
    \tikzstyle{channel} = [fill=white,thin,draw=black,rounded corners=2mm]
    \tikzstyle{cnot} = [draw,fill=white,shape=circle,minimum size=5pt, inner sep=0pt] {+}
    \matrix[row sep=0.4cm, column sep=0.3cm] (circuit) {
    \node (q1) {$\ket{\psi}$}; & 
    \node[operator] (H11) {H}; &
    \node[phase] (P12) {}; &
    \node[phase] (P13) {}; &
    \node[operator] (H14) {H}; &
    \coordinate (mid11); &
    \coordinate (mid12); &
    \node[operator] (H15) {H}; &
    \node[phase] (P16) {}; &
    \node[phase] (P17) {}; &
    \node[cnot] (C18) {+}; &
    \node[operator] (H19) {H}; &
    \node[operator] (Z11) {Z};
    &[-0.3cm]
    \coordinate (end1); \\
    \node (q2) {\ket{1}}; &
    &
    \node[cnot] (C22) {+}; &
    &
    \node[operator] (H24) {H}; &
    \coordinate (mid21); &
    \coordinate (mid22); &
    \node[operator] (H25) {H}; &
    &
    \node[cnot] (C27) {+}; &
    \node[phase] (P28) {}; &
    &
    &
    \coordinate (end2);\\
    \node (q3) {\ket{1}}; &
    &
    &
    \node[cnot] (C33) {+}; &
    \node[operator] (H34) {H}; &
    \coordinate (mid31); &
    \coordinate (mid32); &
    \node[operator] (H35) {H}; &
    \node[cnot] (C36) {+}; &
    &
    \node[phase] (P38) {}; &
    &
    &
    \coordinate (end3); \\
    };
    \begin{pgfonlayer}{background}
        \node[surround] (background) [fit = (q1) (Z11) (H35) (end3)] {};
        \draw[thick] (q1) -- (end1)  (q2) -- (end2) (q3) -- (end3) (P12) -- (C22) (P13) -- (C33) (P16) -- (C36) (P17) -- (C27) (C18) -- (P38);
    \end{pgfonlayer}
    Draw channel box
    \node[channel] (noise) [fit = (mid11) (mid32)] {E\\r\\r\\o\\r};
    \end{tikzpicture}
\caption{Circuit diagram for 3 qubit phase flip repetition code}
\label{undo}
\end{figure}
}

The schematic diagram for undoing a quantum measurement is given in 
Figure~\ref{undo}. This idea~\cite{schindler13} motivated us to devise a novel 
quantum secret sharing protocol based on quantum error correcting codes. It is
resilient in the sense that it survives when the line of Bob or Charlie or 
both is compromised.

\subsection{Our contribution}
In light of the above discussion, let us now explain our exact proposal
and its importance. In line of Figure~\ref{undo}, we also like to refer to
Figure~\ref{qss}. At the bottom of the figure, $\ket{\psi_i}$ below each
column denotes the joint 3-qubit state immediately after the application
of the gates in that column. For example, $\ket{\psi_3}$
is the state after the application of three Hadamard gates in the right half
of the diagram.

{\small
\begin{figure}
\centering
\begin{tikzpicture}[thick]
    \tikzstyle{operator} = [draw,fill=white,minimum size=1.5em] 
    \tikzstyle{phase} = [draw,fill,shape=circle,minimum size=5pt,inner sep=0pt]
    \tikzstyle{surround} = [fill=blue!10,thick,draw=black,rounded corners=2mm]
    \tikzstyle{channel} = [fill=white,thin,draw=black,rounded corners=2mm]
    \tikzstyle{cnot} = [draw,fill=white,shape=circle,minimum size=5pt, inner sep=0pt] {+}
    \matrix[row sep=0.4cm, column sep=0.3cm] (circuit) {
    \node (q1) {$\ket{\psi}$}; & 
    \node[operator] (H11) {H}; &
    \node[phase] (P12) {}; &
    \node[phase] (P13) {}; &
    \node[operator] (H14) {H}; &
    \coordinate (mid11); &
    \coordinate (mid12); &
    \node[operator] (H15) {H}; &
    \node[phase] (P16) {}; &
    \node[phase] (P17) {}; &
    \node[cnot] (C18) {+}; &
    \node[operator] (H19) {H}; &
    \node[operator] (Z11) {Z};
    &[-0.3cm]
    \coordinate (end1); \\
    \node (q2) {\ket{1}}; &
    &
    \node[cnot] (C22) {+}; &
    &
    \node[operator] (H24) {H}; &
    \coordinate (mid21); &
    \coordinate (mid22); &
    \node[operator] (H25) {H}; &
    &
    \node[cnot] (C27) {+}; &
    \node[phase] (P28) {}; &
    &
    &
    \coordinate (end2);\\
    \node (q3) {\ket{1}}; &
    &
    &
    \node[cnot] (C33) {+}; &
    \node[operator] (H34) {H}; &
    \coordinate (mid31); &
    \coordinate (mid32); &
    \node[operator] (H35) {H}; &
    \node[cnot] (C36) {+}; &
    &
    \node[phase] (P38) {}; &
    &
    &
    \coordinate (end3); \\
    \node {$\ket{\psi_0}$}; &
    \node {$\ket{\psi_1}$}; &
    &
    \node {$\ket{\psi_2}$}; &
    \node {$\ket{\psi_3}$}; &
    &
    &
    \node {$\ket{\psi_4}$}; &
    &
    \node {$\ket{\psi_5}$}; &
    \node {$\ket{\psi_{6}}$}; &
    \node {$\ket{\psi_{7}}$}; &
    \node {$\ket{\psi_{8}}$}; &
    \coordinate (end4); \\
    };
    \begin{pgfonlayer}{background}
        \node[surround] (background) [fit = (q1) (Z11) (H35) (end3)] {};
        \draw[thick] (q1) -- (end1)  (q2) -- (end2) (q3) -- (end3) (P12) -- (C22) (P13) -- (C33) (P16) -- (C36) (P17) -- (C27) (C18) -- (P38);
    \end{pgfonlayer}
    Draw channel box
    \node[channel] (noise) [fit = (mid11) (mid32)] {E\\a\\v\\e\\s};
    \end{tikzpicture}
\caption{Schematic diagram for our quantum secret sharing scheme}
\label{qss}
\end{figure}
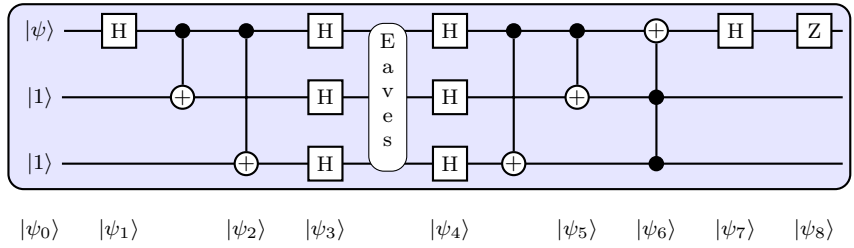
}

The single qubit state $\ket{\psi}$, as in Figure~\ref{qss}, is to be 
distributed between three parties Alice, Bob and Charlie. Two additional 
$\ket{1}$ qubits are used by the distributor for this purpose. Each
individual particle of the three qubit state $\ket{\psi_3}$ is distributed
to Alice, Bob and Charlie. Without loss of generalization, it is considered  
that when they come together to generate the secret, then Bob and Charlie 
will co-operate with Alice and the secret will be generated with Alice.
The particle with Alice corresponding to the three-qubit state $\ket{\psi_8}$
should be same as $\ket{\psi}$ for a successful execution of the protocol.

We claim that our model is resilient. We consider that Bob or Charlie
may cheat. This is pointed out by marking ``Eaves" instead in the figure.
As an example, by cheating, we mean that Bob or Charlie
may measure their individual qubits in certain basis to gain some information
about the secret state $\ket{\psi}$. In our protocol, Alice can detect 
cheating with certain probabilities if either Bob or Charlie or both 
Bob and Charlie cheat(s). Further, the cheater will not be able to obtain any 
information about the secret state from the particle associated with him.

One can consider that parties other than Alice, Bob and Charlie may also be
interested to gain information about the secret to disturb the protocol. 
However, it is enough to consider that Bob and Charlie themselves may play
that adversarial role. We consider that Alice will be honest and the secret
will be prepared by her in cooperation with Bob and Charlie.

Let us present an example. Consider that the cheating is done by Bob and/or
Charlie and they measure in $\{\ket{0}, \ket{1}\}$ basis.
Alice will not disturb the particle that is with her, but
in case of cheating, she has to operate a phase gate to get back the secret. 
On the other hand, Alice can always measure the other two particles 
corresponding to Bob and Charlie. Thus, Alice can measure those two qubits in 
$\{\ket{00}, \ket{01}, \ket{10}, \ket{11}\}$ basis to obtain relevant
information regarding cheating. 

To the best of our knowledge, for quantum secret sharing, there exists no 
such scheme in literature that can decide 
\begin{itemize}
\item whether the cheating is done,
\item whether the expected state is generated or not and
\item whether the exact state could be achieved after some transformation.
\end{itemize}

\section{The Proposed Secret Sharing Protocol}
\label{prot}
First, the dealer $D$ encodes the secret using the encoding circuit (left part 
of Figure~\ref{qss}, up to $\ket{\psi_3}$). The dealer then distributes all 
three qubits among the participants: Alice, Bob and Charlie. Each of them 
applies Hadamard gate after getting the qubits. Now, one of them, say Alice, 
decides to generate the secret and so the other two delivers their qubits 
to Alice.

Alice applies two CNOT gates (as shown in Fig~\ref{qss}) considering 
her qubit as the control. Next, she applies Toffoli gates considering
Bob and Charlie's bits as control. Then she applies Hadamard (H) and 
phase (Z) gates to generate the secret at her end.
Alice measures the particles corresponding to Bob and Charlie in 
$\{\ket{00}, \ket{01}, \ket{10}, \ket{11}\}$ basis. 
\begin{itemize}
\item If she gets $\ket{11}$, she decides that no cheating has been done in 
the process. 
\item If she gets $\ket{01}$, she understands that cheating is done 
by Bob. 
\item If she gets $\ket{10}$, she understands that cheating is performed by 
Charlie. 
\item If she gets $\ket{00}$, she knows that both Bob and Charlie have
cheated. 
\end{itemize}
In all three cases related to cheating, she has to apply a phase gate 
to generate the secret. On the other hand we have the following cases related
to cheating.
\begin{itemize}
\item If Bob has cheated by measuring in $\{\ket{0}, \ket{1}\}$ basis, then
Alice will observe either $\ket{01}$ or $\ket{11}$ 
with probability $\frac{1}{2}$ each.
\item If Charlie has cheated by measuring in $\{\ket{0}, \ket{1}\}$ basis, then
Alice will observe either $\ket{10}$ or $\ket{11}$
with probability $\frac{1}{2}$ each.
\item If both Bob and Charlie have cheated by measuring their particle
in $\{\ket{0}, \ket{1}\}$ basis individually, then
Alice will observe any of $\ket{00}, \ket{01}, \ket{10}, \ket{11}$  
with probability $\frac{1}{4}$ each.
\end{itemize}
Thus, in any case, the secret quantum state will be created perfectly. However,
Alice will be able to understand whether the cheating has been done with 
certain probabilities. 

\subsection{Detailed Calculation}
We show the step-by-step evolution of the joint state as the protocol runs
from left to right in Figure~\ref{qss}.
We can write 
$$\ket{\psi_0} = \left(\alpha\ket{0} + \beta\ket{1}\right)\ket{1}\ket{1}.$$
After the application of the Hadamard gate, the joint state becomes
$$\ket{\psi_1} = \frac{1}{\sqrt{2}}\left[\alpha\left(\ket{0} + 
\ket{1}\right) +\beta\left(\ket{0}-\ket{1}\right)\right]\ket{1}\ket{1}.$$
After the application of the two CNOT gates, the state becomes
$$\ket{\psi_2} = \frac{1}{\sqrt{2}}\left[\alpha\left(\ket{011} + 
\ket{100}\right) +\beta\left(\ket{011}-\ket{100}\right)\right].$$
Next, the dealer applies three Hadamard gates, one on each line, yielding
the state
\begin{equation}
\label{psi3}
\ket{\psi_3} = \frac{1}{2}\left[
\alpha\left(\ket{000} - \ket{110} + \ket{011} - \ket{101}\right) +
\beta\left(\ket{100}-\ket{010} - \ket{001} + \ket{111}\right)\right].
\end{equation}
Now, the dealer distributes this state amongst Alice, Bob and Charlie,
one particle to each.
After getting the qubits, each party applies Hadamard on his/her qubit.
So the resultant state becomes 
$$\ket{\psi_4} = \frac{1}{\sqrt{2}}\left[\alpha\left(\ket{011} + 
\ket{100}\right) +\beta\left(\ket{011}-\ket{100}\right)\right] = \ket{\psi_2}.$$
After two CNOT gates, the state becomes
$$\ket{\psi_5} = \frac{1}{\sqrt{2}}\left[\alpha\left(\ket{011} + 
\ket{111}\right) +\beta\left(\ket{011}-\ket{111}\right)\right].$$
Next, after application of the Toffoli gate, the state becomes 
$$\ket{\psi_6} = \frac{1}{\sqrt{2}}\left[\alpha\left(\ket{0} + 
\ket{1}\right) - \beta\left(\ket{0}-\ket{1}\right)\right]\ket{1}\ket{1}.$$
After the last Hadamard gate, it reduces to
$$\ket{\psi_7} = \left(\alpha\ket{0} - \beta\ket{1}\right)\ket{1}\ket{1}.$$
The phase gate in the end converts this state into
$$\ket{\psi_8} = \left(\alpha\ket{0} + \beta\ket{1}\right)\ket{1}\ket{1} = \ket{\psi_0}.$$
Now Alice measures her ancilla in the 
$\{\ket{00}, \ket{01}, \ket{10}, \ket{11}\}$ basis. 
In case she gets $\ket{11}$, she knows that the secret has been successfully
generated.

\section{Resiliency of the scheme and analysis of cheating}
In this section, we analyze what happens when Bob or Charlie or both 
tries to cheat to the corresponding lines. We do not consider any other
eavesdropper as any cheating by other parties can also be done by 
Bob and Charlie themselves.

\subsection{When Bob attempts cheating}
As Bob wants to get some information about the secret, the way we consider
is to measure the qubit just after he receives it. Assuming that he measures in
the computational basis, he gets $\ket{0}$ with probability $\frac{1}{2}$
and $\ket{1}$ with probability $\frac{1}{2}$. 

We have the state $\ket{\psi_3}$ in hand.
Suppose Bob gets $\ket{0}$. From Equation~\eqref{psi3}, we can write the 
remaining two-qubit state as
$$\ket{\psi_3^{AC}} = \frac{1}{\sqrt{2}}\left[
\alpha\left(\ket{00} - \ket{11}\right) +
\beta\left(\ket{10} - \ket{01}\right)\right].$$
After the measurement by Bob, he again replaces the measured qubit 
(here $\ket{0}$). Then each of Alice, Bob and Charlie applies Hadamard, and 
the joint state becomes $$\ket{\psi^{AB'_0C}_4} = \frac{1}{2}\left[
\alpha\left(\ket{001} + \ket{100} + \ket{110} + \ket{011}\right) +
\beta\left(\ket{001}-\ket{100} - \ket{110} + \ket{011}\right)\right].$$
Now, Bob and Charlie give their qubits to Alice. Alice applies CONTs considering
her qubit as control. Then the overall state becomes
$$\ket{\psi^{AB'_0C}_5} = \frac{1}{2}\left[
\alpha\left(\ket{001} + \ket{111} + \ket{101} + \ket{011}\right) +
\beta\left(\ket{001}-\ket{111} - \ket{101} + \ket{011}\right)\right].$$
Next, Alice operates Toffoli considering Bob and Charlie's qubits as control.
Then we get
$$\ket{\psi^{AB'_0C}_6} = \frac{1}{2}\left[
\alpha\left(\ket{0} + \ket{1}\right)\left(\ket{01} + \ket{11}\right) +
\beta\left(\ket{0}-\ket{1}\right)\left(\ket{01} - \ket{11}\right)\right].$$
Now, Alice applies Hadamard on the first qubit. Thus, the resultant state
becomes
$$\ket{\psi^{AB'_0C}_7} = \frac{1}{\sqrt{2}}\left[
\left(\alpha\ket{0} - \beta\ket{1}\right)\ket{11} +
\left(\alpha\ket{0} + \beta\ket{1}\right)\ket{01}\right].$$
After Alice applies the last phase gate, the state becomes
\begin{equation}
\label{psiab0c}
\ket{\psi^{AB'_0C}_8} = \frac{1}{\sqrt{2}}\left[
\left(\alpha\ket{0} + \beta\ket{1}\right)\ket{11} +
\left(\alpha\ket{0} - \beta\ket{1}\right)\ket{01}\right].
\end{equation}

Now, Alice measures the particles corresponding to Bob and Charlie in 
the basis $\{\ket{00}, \ket{01}, \ket{10}, \ket{11}\}$. In this case, she gets
$\ket{11}$ with probability $\frac{1}{2}$ and $\ket{01}$ with probability
$\frac{1}{2}$. In the first case, she gets back the secret, but cannot detect
that Bob has cheated. In the second case, she discovers that
Bob has indeed cheated and she can get back the secret by
applying one more phase gate to her qubit. 

Similarly, when Bob measures $\ket{1}$, the final state becomes
\begin{equation}
\label{psiab1c}
\ket{\psi^{AB'_1C}_8} = \frac{1}{\sqrt{2}}\left[
\left(\alpha\ket{0} + \beta\ket{1}\right)\ket{11} -
\left(\alpha\ket{0} - \beta\ket{1}\right)\ket{01}\right].
\end{equation}
As can be seen by comparing Equations~\eqref{psiab0c} and~\eqref{psiab1c},
we get the same result after Alice measures the particles corresponding
to Bob and Charlie in computational basis. 

\subsection{When Charlie attempts cheating}
In a similar manner, Charlie measures his qubit just after he receives it. 
Assuming that he measures in the computational basis, he gets $\ket{0}$ with 
probability $\frac{1}{2}$ and $\ket{1}$ with probability $\frac{1}{2}$. 

Suppose Charlie obtains $\ket{0}$. From Equation~\eqref{psi3}, we can 
write the remaining two-qubit state as
$$\ket{\psi_3^{AB}} = \frac{1}{\sqrt{2}}\left[
\alpha\left(\ket{00} - \ket{11}\right) +
\beta\left(\ket{10} - \ket{01}\right)\right].$$
Now, each applies Hadamard on the respective input and the joint state becomes
$$\ket{\psi^{ABC'_0}_4} = \frac{1}{2}\left[
\alpha\left(\ket{100} + \ket{010} + \ket{011} + \ket{101}\right) -
\beta\left(\ket{100}-\ket{010} - \ket{011} + \ket{101}\right)\right].$$
Now, Bob and Charlie give their qubits to Alice. Alice applies CONTs 
considering her qubit as control. Then she obtains the overall state as
$$\ket{\psi^{ABC'_0}_5} = \frac{1}{2}\left[
\alpha\left(\ket{111} + \ket{010} + \ket{011} + \ket{110}\right) -
\beta\left(\ket{111}-\ket{010} - \ket{011} + \ket{110}\right)\right].$$
Next, Alice operates Toffoli considering Bob and Charlie's qubits as control.
Then we get
$$\ket{\psi^{ABC'_0}_6} = \frac{1}{2}\left[
\alpha\left(\ket{0} + \ket{1}\right)\left(\ket{11} + \ket{10}\right) -
\beta\left(\ket{0}-\ket{1}\right)\left(\ket{11} - \ket{10}\right)\right].$$
Now, Alice applies Hadamard on the first qubit. Thus, the resultant state
becomes
$$\ket{\psi^{ABC'_0}_7} = \frac{1}{\sqrt{2}}\left[
\left(\alpha\ket{0} - \beta\ket{1}\right)\ket{11} +
\left(\alpha\ket{0} + \beta\ket{1}\right)\ket{10}\right].$$
After Alice applies the last phase gate, the state becomes
\begin{equation}
\label{psiabc0}
\ket{\psi^{ABC'_0}_8} = \frac{1}{\sqrt{2}}\left[
\left(\alpha\ket{0} + \beta\ket{1}\right)\ket{11} +
\left(\alpha\ket{0} - \beta\ket{1}\right)\ket{10}\right].
\end{equation}
Alice measures the qubits of Bob and Charlie in 
$\{\ket{00}, \ket{01}, \ket{10}, \ket{11}\}$ basis. In this case, she gets
$\ket{11}$ with probability $\frac{1}{2}$ and $\ket{10}$ with probability
$\frac{1}{2}$. In the first case, she gets back the secret, but cannot detect
that Charlie tried to cheat. In the second case, she knows for certain that
Charlie has indeed cheated and she can get back the secret simply by
applying one more phase gate to her qubit. 

Similarly, when Charlie measures $\ket{1}$, the final state becomes
\begin{equation}
\label{psiabc1}
\ket{\psi^{ABC'_1}_8} = \frac{1}{\sqrt{2}}\left[
\left(\alpha\ket{0} + \beta\ket{1}\right)\ket{11} -
\left(\alpha\ket{0} - \beta\ket{1}\right)\ket{10}\right].
\end{equation}
As can be seen by comparing Equations~\eqref{psiabc0} and~\eqref{psiabc1},
we get the same result. 

\subsection{When both Bob and Charlie are Eavesdroppers}
Assume that both Bob and Charlie measure in the computational basis.
Suppose, Bob gets $\ket{0}$ and Charlie also gets $\ket{0}$. 
From Equation~\eqref{psi3}, we can write the 
remaining state as
$$\ket{\psi_3^{A}} = \alpha\ket{0} + \beta\ket{1}.$$
After each applies Hadamard, the joint state becomes
\begin{eqnarray*}
\ket{\psi^{AB'_0C'_0}_4} & = & \frac{1}{2\sqrt{2}}\left[
\alpha\left(\ket{000} + \ket{001} + \ket{010} + \ket{011} + 
\ket{100} + \ket{101} + \ket{110} + \ket{111}\right) \right.\\
& & \left. + \beta\left(\ket{000} + \ket{001} + \ket{010} + \ket{011} - 
\ket{100} - \ket{101} - \ket{110} - \ket{111}\right)\right].
\end{eqnarray*}
Now, Bob and Charlie give their qubits to Alice. Alice applies CONTs considering
her qubit as control. It can be easily verified that after this, 
the overall state remains the same, i.e.,
$\ket{\psi^{AB'_0C'_0}_5} = \ket{\psi^{AB'_0C'_0}_4}$.

Next, Alice operates Toffoli considering Bob and Charlie's qubits as control.
Then we get the resultant state as
\begin{eqnarray*}
\ket{\psi^{AB'_0C'_0}_6} & = & \frac{1}{2\sqrt{2}}\left[
\alpha\left(\ket{000} + \ket{001} + \ket{010} + \ket{011} + 
\ket{100} + \ket{101} + \ket{110} + \ket{111}\right) \right.\\
& & \left. + \beta\left(\ket{000} + \ket{001} + \ket{010} - \ket{011} - 
\ket{100} - \ket{101} - \ket{110} + \ket{111}\right)\right].
\end{eqnarray*}
Now, Alice applies Hadamard on the first qubit. Thus, the resultant state
becomes
$$\ket{\psi^{AB'_0C'_0}_7} = \frac{1}{\sqrt{2}}\left[
\left(\alpha\ket{0} - \beta\ket{1}\right)\ket{11} +
\left(\alpha\ket{0} + \beta\ket{1}\right)
\left(\ket{00} + \ket{01} + \ket{10}\right)\right].$$
After Alice applies the last phase gate, the state becomes
$$\ket{\psi^{AB'_0C'_0}_7} = \frac{1}{\sqrt{2}}\left[
\left(\alpha\ket{0} + \beta\ket{1}\right)\ket{11} +
\left(\alpha\ket{0} - \beta\ket{1}\right)
\left(\ket{00} + \ket{01} + \ket{10}\right)\right].$$
Now, Alice measures the qubits of Bob and Charlie in 
$\{\ket{00}, \ket{01}, \ket{10}, \ket{11}\}$ basis. In this case, she gets
each basis state with probability $\frac{1}{4}$. The cases can be summarized
as follows.
\begin{itemize}
\item If she gets $\ket{11}$, she gets back the secret, but cannot detect
if anyone tried to eavesdrop.
\item If she gets $\ket{01}$, she knows that Bob was eavesdropping.
\item If she gets $\ket{10}$, she knows that Charlie was eavesdropping.
\item If she gets $\ket{00}$, she knows that both Bob and Charlie were 
eavesdropping.
\end{itemize}

By similar analysis of the cases when Bob and Charlie measure the other 
three possible pairs of outputs, we get the same result.

\section{Analysis of Arbitrary Measurements}
The protocol that we discussed can withstand errors introduced due to 
measurement by the advisory in the computational basis. If the advisory
performs the measurement in other basis, then the secret may not be 
recovered. In this section we first analyse what happens when the adversary
measures in arbitrary basis and then discuss how our secret sharing protocol
can be modified to tolerate measurement in any basis. 

\subsection{Analysis of Measurement in Arbitrary Basis}
Let Bob (without loss of generality) measure in 
$\{\ket{\gamma},\ket{\gamma^{\perp}}\}$ basis, where
$\ket{\gamma} = a\ket{0} + b\ket{1}$ and
$\ket{\gamma^{\perp}} = b^{*}\ket{0} - a^{*}\ket{1}$, 
with $a, b \in \mathbb{C}$ and $|a|^2 + |b|^2 = 1$. Then we can write 
$\ket{0} = a^{*}\ket{\gamma} + b\ket{\gamma^{\perp}}$
and $\ket{1} = b^{*}\ket{\gamma} - a\ket{\gamma^{\perp}}$.

Thus, we can rewrite Equation~\eqref{psi3} as 
\begin{eqnarray*}
\ket{\psi_3} & = & \frac{1}{2}\left[
\alpha\left(\ket{000} - \ket{110} + \ket{011} - \ket{101}\right) +
\beta\left(\ket{100}-\ket{010} - \ket{001} + \ket{111}\right)\right]\\
& = & \frac{1}{2}\left[
\alpha\left(\ket{0}(a^{*}\ket{\gamma} + b\ket{\gamma^{\perp}})\ket{0} - \ket{1}(b^{*}\ket{\gamma} - a\ket{\gamma^{\perp}})\ket{0} \right.\right.\\
& & + \left.\ket{0}(b^{*}\ket{\gamma} - a\ket{\gamma^{\perp}})\ket{1} - 
\ket{1}(a^{*}\ket{\gamma} + b\ket{\gamma^{\perp}})\ket{1}\right) \\
& & + \beta\left(\ket{1}(a^{*}\ket{\gamma} + b\ket{\gamma^{\perp}}){\ket0}-\ket{0}(b^{*}\ket{\gamma} - a\ket{\gamma^{\perp}})\ket{0}\right. \\
& & - \left.\left.\ket{0}(a^{*}\ket{\gamma} + b\ket{\gamma^{\perp}})\ket{1} + 
\ket{1}(b^{*}\ket{\gamma} - a\ket{\gamma^{\perp}})\ket{1}\right)\right].
\end{eqnarray*}

When Bob measures this state in $\{\ket{\gamma},\ket{\gamma^{\perp}}\}$ basis, 
it can be easily seen that she gets each of $\ket{\gamma}$ and 
$\ket{\gamma^{\perp}}$ with probability $\frac{1}{2}$.

Suppose Bob observes $\ket{\gamma}$. The remaining state becomes
\begin{eqnarray*}
\ket{\psi_3^{AC}} & = & \frac{1}{\sqrt{2}}\left[
\alpha\left(a^{*}\ket{00} - b^{*}\ket{10} + b^{*}\ket{01} - a^{*}\ket{11}\right)\right.\\
& & + \left.\beta\left(a^{*}\ket{10} - b^{*}\ket{00} - a^{*}\ket{01} + b^{*}\ket{11}\right)\right].
\end{eqnarray*}

After each applies Hadamard, the joint state becomes
\begin{eqnarray*}
\ket{\psi^{AB'_\gamma C}_4} & = & 
\frac{1}{2}\left[
(\alpha a^{*} - \beta b^{*} - \alpha b^{*} + \beta a^{*})(a+b)\ket{001}\right.\\
& & + (\alpha a^{*} - \beta b^{*} + \alpha b^{*} - \beta a^{*})(a+b)\ket{100}\\
& & + (\alpha a^{*} - \beta b^{*} + \alpha b^{*} - \beta a^{*})(a-b)\ket{110}\\
& & + \left. (\alpha a^{*} - \beta b^{*} - \alpha b^{*} + \beta a^{*})(a-b)\ket{011}
\right].
\end{eqnarray*}
Now, Bob and Charlie give their qubits to Alice. Alice applies CONTs considering
her qubit as control. Then the overall state becomes
\begin{eqnarray*}
\ket{\psi^{AB'_\gamma C}_5} & = & 
\frac{1}{2}\left[
(\alpha a^{*} - \beta b^{*} - \alpha b^{*} + \beta a^{*})(a+b)\ket{001}\right.\\
& & + (\alpha a^{*} - \beta b^{*} + \alpha b^{*} - \beta a^{*})(a+b)\ket{111}\\
& & + (\alpha a^{*} - \beta b^{*} + \alpha b^{*} - \beta a^{*})(a-b)\ket{101}\\
& & + \left. (\alpha a^{*} - \beta b^{*} - \alpha b^{*} + \beta a^{*})(a-b)\ket{011}
\right].
\end{eqnarray*}

Next, Alice operates Toffoli considering Bob and Charlie's qubits as control.
Then we get
\begin{eqnarray*}
\ket{\psi^{AB'_\gamma C}_6} & = & 
\frac{1}{2}\left[
(\alpha a^{*} - \beta b^{*} - \alpha b^{*} + \beta a^{*})(a+b)\ket{001}\right.\\
& & + (\alpha a^{*} - \beta b^{*} + \alpha b^{*} - \beta a^{*})(a+b)\ket{011}\\
& & + (\alpha a^{*} - \beta b^{*} + \alpha b^{*} - \beta a^{*})(a-b)\ket{101}\\
& & + \left. (\alpha a^{*} - \beta b^{*} - \alpha b^{*} + \beta a^{*})(a-b)\ket{111}
\right].
\end{eqnarray*}
Now, Alice applies Hadamard on the first qubit. Thus, the resultant state
becomes
\begin{eqnarray*}
\ket{\psi^{AB'_\gamma C}_7} & = & \frac{1}{\sqrt{2}}\left[
\left(\left(\alpha - \beta(ab^*+a^*b)\right)\ket{0}
+ \left(\alpha(ab^*+a^*b) - \beta\right)\ket{1}\right)\ket{11} \right. \\
& & + \left(\left(\alpha(|a|^2-|b|^2) + \beta(a^*b-ab^*)\right)\ket{0}\right.\\
& & \left.\left. - \left(\alpha(ab^*-a^*b) + \beta(|b|^2-|a|^2)\right)\ket{1}\right)\ket{01} \right].
\end{eqnarray*}
After Alice applies the last phase gate, the state becomes
\begin{eqnarray}
\label{psiabgammac}
\ket{\psi^{AB'_\gamma C}_8} & = & \frac{1}{\sqrt{2}}\left[
\left(\left(\alpha - \beta(ab^*+a^*b)\right)\ket{0}
- \left(\alpha(ab^*+a^*b) - \beta\right)\ket{1}\right)\ket{11} \right. \nonumber \\
& & + \left(\left(\alpha(|a|^2-|b|^2) + \beta(a^*b-ab^*)\right)\ket{0}\right.\nonumber \\
& & \left.\left. + \left(\alpha(ab^*-a^*b) + \beta(|b|^2-|a|^2)\right)\ket{1}\right)\ket{01} \right].
\end{eqnarray}

Note that due to the measurement, $\alpha$ is shifted to
$\alpha' = \alpha - \beta(ab^*+a^*b)$ and $\beta$ is shifted to
$\beta' = \beta - \alpha(ab^*+a^*b)$. Thus, depending on the values 
of $a$ and $b$, the secret evolves.

Similar analysis holds for the case when Bob measures $\ket{\gamma^{\perp}}$.
Continuing with the analysis of Bob's measurement as $\ket{\gamma}$, we find
that when Alice measures the qubits corresponding to Bob and Charlie in 
$\{\ket{00}, \ket{01}, \ket{10}, \ket{11}\}$ basis, she gets
$\ket{11}$ with probability $\frac{1}{2}$ and $\ket{01}$ with probability
$\frac{1}{2}$. Note that if one takes $a = 1, b = 0$, then 
Equation~\eqref{psiabgammac} coincides with Equation~\eqref{psiab0c}. 
On the other hand, if one takes $a = 0, b = 1$, then 
Equation~\eqref{psiabgammac} coincides with Equation~\eqref{psiab1c}.

Next, let us consider the case $a = b = \frac{1}{\sqrt{2}}$, i.e., the
measurement basis is $\{\ket{+}, \ket{-}\}$. In this case, when Bob measures
$\ket{+}$, Equation~\eqref{psiabgammac} reduced to
$$\ket{\psi^{AB'_+C}_8} = \frac{1}{\sqrt{2}}(\alpha - \beta)\left(\ket{0} - \ket{1}\right)\ket{11}.$$
Upon measuring her ancilla bits in the computational basis, Alice would
get wrong information that there is no eavesdropping. Moreover, the state
is also not recovered properly at Alice's end. Similar cases can be analysed 
considering Charlie or both. 

Thus, if the participants go for a measurement in arbitrary basis, then it 
is not possible to reconstruct the secret with simple quantum error correcting 
codes. However, with codes that can correct any kind of error, it is possible
to design quantum secret sharing schemes that are more resilient. Below 
we outline this extension. 

\subsection{Possible Resiliency against Arbitrary Operations by Cheaters}
In this section, we discuss remedies against not only arbitrary measurements
but also arbitrary interactions by the Cheaters.
The protocol described in Section~\ref{prot} uses three qubit phase flip 
repetition code. Here, we modify the protocol with 
nine qubit Shor code~\cite{shor95}. This code is a combination of three qubit 
phase flip and bit flip codes. The qubit is first encoded using the 
phase flip code, i.e., $\ket{0}$ is encoded as $\ket{+++}$ and
$\ket{1}$ is represented as $\ket{---}$. Next, each of these qubits 
are encoded using three qubit bit flip code, i.e., $\ket{+}$ is encoded as
$\ket{000}+\ket{111})/\sqrt{2}$ and $\ket{-}$ is encoded as
$\ket{000}-\ket{111})/\sqrt{2}$.

Suppose, before the interaction by the adversary, the state of the system
is $\ket{\psi_{ABC}}$. Any error $\Upsilon$, including the one due to 
interaction, can be analyzed by expanding it in an operator-sum
representation with operation elements $\{E_i\}$, we would have the
post-interaction state as 
$$\Upsilon \ket{\psi_{ABC}}\bra{\psi_{ABC}} 
= \sum_i E_i \ket{\psi_{ABC}}\bra{\psi_{ABC}} E_i^{\dagger}.$$
As an operator, $E_i$ can be expanded as a linear combination of
the identity, the bit flip $X$, the phase flip $Z$ and the combined 
bit and phase flip $XZ$. Measuring the error syndrome would collapse 
this superposition into one of the four states, namely,
$\ket{\psi_{ABC}}$, $X\ket{\psi_{ABC}}$, $Z\ket{\psi_{ABC}}$ and
$XZ\ket{\psi_{ABC}}$. By applying the appropriate inverse operation, the
recovery is possible.

\section{Conclusion}
In this paper, we show how quantum error correcting codes can be suitably
used to present quantum secret sharing schemes with certain kind of resiliency.
Our work is not in the line of mostly studied quantum secret sharing that lends
the idea from teleportation.  
We explain the basic idea with a simple code and point out its merits and
demerits. We also outline how any arbitrary adversarial attempt can be resisted
with more complicated quantum error correcting codes. Detailed study with
different kinds of access structures and finding out suitable error correcting
codes are important as future works in this direction.


\begin{thebibliography}{}

\bibitem{AP06}
P. Agrawal, A. Pati.
Perfect teleportation and superdense coding with W states.
{\it Phys. Rev. A} {\bf 74}, 062320 (2006).

\bibitem{BB93}
C. H. Bennett, G. Brassard, C. Cr\'epeau, R. Jozsa, A. Peres, W. K. Wootters.
Teleporting an unknown quantum state via dual classical and Einstein-Podolsky-Rosen channels. {\it Phys. Rev. Lett.} {\bf 70} (1993), pp.~1895--1899.

\bibitem{BR01}
H. J. Briegel, R. Raussendorf.
Persistent Entanglement in Arrays of Interacting Particles.
{\it Phys. Rev. Lett.} {\bf 86}, 910--913 (2001).

\bibitem{C99}
R. Cleve, D. Gottesman, H.K. Lo.
How to Share a Quantum Secret.
{\it Phys. Rev. Lett.} {\bf 83}, 648--651 (1999).

\bibitem{D05}
F.G. Deng, C. Li, Y. Li, P. Zhou, H. Zhou.
Multiparty quantum-state sharing of an arbitrary two-particle state with Einstein-Podolsky-Rosen pairs.
{\it Physics Rev. A} {\bf 72}, 044301 (2005).

\bibitem{H12}
L.F. Han, H. F. Xu. 
Probabilistic and Controlled Teleportation of an Arbitrary Two-Qubit State 
via One Dimensional Five-Qubit Cluster-Class State.
{\it Int. J. Theor. Phys.} {\bf 51}, 2540--2554 (2012).

\bibitem{HBB99}
M. Hillery, V. Bu\v{z}ek, A. Berthiaume.
Quantum secret sharing.
{\it Phys. Rev. A} {\bf 59}, 1829--1834 (1999).

\bibitem{KB98}
A. Karlsson, M. Bourennane.
Quantum teleportation using three-particle entanglement.
{\it Phys. Rev. A} {\bf 58}, 4394--4400 (1998).

\bibitem{KK99}
A. Karlsson, M. Koashi, N. Imoto.
Quantum entanglement for secret sharing and secret splitting.
{\it Phys. Rev. A} {\bf 59}, 162--168 (1999).

\bibitem{NJ10}
Y. Li, J. Liu, Y. Nie.
Quantum Teleportation and Quantum Information Splitting by Using a Genuinely Entangled Six-Qubit State.
{\it Int. J. Theor. Phys.} {\bf 49}, 2592--2599 (2010).

\bibitem{NL10}
J. Liu, Y. Li, Y. Nie.
Controlled Teleportation of an Arbitrary Two-Particle Pure or Mixed State by Using a Five-Qubit Cluster State.
{\it Int. J. Theor. Phys.} {\bf 49}, 1976--1984 (2010).

\bibitem{P08}
S. Muralidharan, P. K. Panigrahi.
Perfect teleportation, quantum-state sharing, and superdense coding through a genuinely entangled five-qubit state.
{\it Phys. Rev. A} {\bf 77}, 032321 (2008).

\bibitem{PS11}
S. Muralidharan, S. Jain, P.K. Panigrahi.
Splitting of quantum information using N-qubit linear cluster states.
{\it Opt. Commun.} {\bf 284}, 1082--1085 (2011).

\bibitem{NS11}
Y. Nie, M. Sang, Y. Li, J. Liu.
Three-Party Quantum Information Splitting of an Arbitrary Two-Qubit State by Using Six-Qubit Cluster State.
{\it Int. J. Theor. Phys.} {\bf 50}, 1367--1371 (2011).

\bibitem{nc04}
M. A. Nielsen and I. L. Chuang.
{\it Quantum Computation and Quantum Information},
Cambridge University Press, 2004.

\bibitem{P00} 
A. K. Pati.
Assisted cloning and orthogonal complementing of an unknown state.
{\it Phys. Rev. A} {\bf 61}, 022308 (2000).

\bibitem{PM11}
N. Paul, J.V. Menon, S. Karumanchi, S. Muralidharan, P.K. Panigrahi.
Quantum tasks using six qubit cluster states.
{\it Quantum Inf. Process.} {\bf 10}, 619--632 (2011).

\bibitem{rietjens05}
K. Rietjens, B. Schoenmakers, P. Tuyls.
Quantum Information Theoretical Analysis of Various Constructions for Quantum 
Secret Sharing.
arXiv:quant-ph/0502009 (2005).

\bibitem{R05}
G. Rigolin.
Quantum teleportation of an arbitrary two-qubit state and its relation to multipartite entanglement.
{\it Phys. Rev. A} {\bf 71}, 032303 (2005).

\bibitem{schindler13}
P. Schindler, T. Monz, D. Nigg, J. T. Barreiro, E. A. Martinez, M. F. Brandl, 
M. Chwalla, M. Hennrich, R. Blatt.
Undoing a Quantum Measurement.
{\it Physical Review Letters}, {\bf 110}, 070403 (2013).

\bibitem{shor95}
P. W. Shor.
Scheme for reducing decoherence in quantum computer memory.
{\it Physical Review A}, {\bf 52} (4), R2493--R2496 (1995).

\bibitem{YC06}
Y. Yeo, W. K. Chua.
Teleportation and Dense Coding with Genuine Multipartite Entanglement.
{\it Phys. Rev. Lett.} {\bf 96}, 060502 (2006).

\bibitem{zhang11}
Z. Zhang, and W. Liu and C. Li.
Quantum secret sharing based on quantum error-correcting codes.
{\it Chinese Physics B} {\bf 20} (5), 050309 (2011). 

\bibitem{Z12}
Q. Y. Zhang, Y. B. Zhan.
Quantum Information Splitting by Using Asymmetric Multi-particle State.
{\it Int. J. Theor. Phys.} {\bf 51}, 3037--3044 (2012).

\bibitem{ZL11}
M. Zhao, Z. Li, S. Fei, Z. Wang, X. Jost.
Faithful teleportation with arbitrary pure or mixed resource states.
{\it J. Phys. A: Math. Theor.} {\bf 44}, 215302 (2011).

\bibitem{Z06}
S.B. Zheng.
Splitting quantum information via W states.
{\it Phys. Rev. A} {\bf 74}, 054303 (2006).

\end{thebibliography}
\end{document}